# Accessible Light Bullets via synergetic nonlinearities

Ian B. Burgess[1,2], Marco Peccianti[1,3*], Gaetano Assanto[4] and Roberto Morandotti[1]

[1] INRS-EMT, Université du Québec, 1650 Blv. L. Boulet, Varennes, Québec J3X 1S2 Canada

[2] School of Eng. Appl. Sci., Harvard University, Pierce Hall, 29 Oxford Street, Cambridge, MA 02138.

[3] Res. Center SOFT INFM-CNR,"Sapienza" University, P. A. Moro 2, 00185 Rome – Italy.

[4] Nonlinear Optics & OptoElectronics Lab, Univ. "Roma Tre", Via Vasca Navale 84, 00146 Rome - Italy.

**Abstract**

We introduce a new form of stable spatio-temporal self-trapped optical packets stemming from the interplay of local and nonlocal nonlinearities. Pulsed self-trapped light beams in media with both electronic and molecular nonlinear responses are addressed to prove that spatial and temporal effects can be decoupled, allowing for independent tuning. We numerically demonstrate that (3+1)D light *bullets* and *anti-bullets*, i. e. bright and dark temporal solitons embedded in stable (2+1)D nonlocal spatial solitons, can be generated in reorientational media under experimentally feasible conditions.

Solitary waves or solitons are ubiquitous in nature and have been identified in a broad range of physical systems, including fluids, plasmas, electromagnetic waves, biologic and atomic matter [1-2]. In optics, temporal and spatial solitons have been recognized as fundamental objects, not only in nonlinear physics, but also for their potential impact in new all-optical signal processing technologies. Optical solitons have been obtained in a broad range of media and, in spite of the diverse underlying nonlinear responses, share several universal properties [3-5]. The main distinguishing feature in this group is the scale of time and space in which the nonlinear mechanisms operate: on one side of this group are those nonlinearities which originate from thermal, molecular, charge drifting mechanisms [6-13], nonlocal in both time and space; on the opposite side are electronic or catalytic nonlinearities, usually treated as local and instantaneous at optical frequencies [14-18]. The instantaneous Kerr effect enables the observation of temporal nonlinear evolution in short optical pulses; in fibers, e. g., several phenomena [self-phase modulation, pulse compression, bright and dark solitons, etc.] have been the subject of numerous studies in literature [14] as they tend to became important when high peak optical power is considered. In the spatial domain (2+1)D solitons, i. e. beams spatially self-trapped in both transverse dimensions, have been shown to be unstable in the Kerr regime [19]. Spatio-temporal solitons, i. e. wave packets that maintain invariant both their spatial and temporal profiles through propagation, sometimes referred as (3+1)D solitons or *bullets*, were first proposed by Silberberg [20] and have since gained much interest, although experimentally feasible conditions for their excitation have yet to be found [21].

Recently, robust and stable (2+1)D spatial solitons were theoretically addressed in media exhibiting a highly nonlocal response [22, 23] and experimentally realized in a variety of materials (liquid crystals, thermal, photorefractive and photochemical media). [6-13] The size of such a nonlinearity is naturally enhanced by a time-integration of the field-dependent response which, in turn, tends to be insensitive to fluctuations faster than a characteristic time.[24] When a local (in time and/or in space) nonlinearity coexists with a nonlocal one, for example the electronic response in thermal/diffusive media, it is possible to tailor a combination of fast/local and slow/nonlocal nonlinearities by separately controlling instantaneous and average excitation properties. This makes available a whole new class of nonlinear dynamical systems where interplay is allowed between phenomena associated to an instantaneous/local response and those linked to a noninstantaneous/nonlocal nonlinearity, extending beyond optics to areas such as Bose-Einstein condensates and plasmas, where multiple nonlinear mechanisms come into play.

In this Letter we predict the existence of a new kind of (3+1)D light *bullets* with spatial and temporal profiles governed by two independent nonlinear processes. This specific synergy can take place in media with both a non-instantaneous nonlinearity and a Kerr-type response.

We numerically demonstrate light *bullets* and *anti-bullets*, i. e. temporal bright and dark solitons within a nonlocal self-localized beam, making specific reference to a reorientational dielectric with a cubic electronic susceptibility, namely Liquid Crystals in the Nematic phase (NLC).[7]

Let us consider a Gaussian light beam composed of a train of pulses, having width and separation much shorter than the characteristic response time of the slow nonlinear mechanism; hence, in a self-focusing slow/nonlocal dielectric, stable and well confined (2+1)D solitons with propagation-invariant profile can be generated as the spatial dynamics are insensitive to the pulsed character of the excitation [6,]. Whereas in principle any slow nonlinear response can be exploited, spatially nonlocal media are specifically addressed here as they provide also the stabilization mechanism required to prevent the Kerr spatial collapse [4,5,22].

In the time domain, each individual pulse is subject to a spatially invariant confining potential with constant (temporal) dispersion for the soliton mode. For the space-time field distribution of a pulse-train with central frequency $\omega_0$, propagating with a wavevector $\boldsymbol{\beta}=\beta\,\mathbf{z}$ along $z$ and with a pulse-to-pulse separation $\sigma$, we can write: $E(\mathbf{r},T) = F(x,y)\{\Sigma_n [A(z, T - n\sigma)] \text{Exp}[i(\beta z - \omega_0 t)]\}$, with $\mathbf{r} = (x, y, z)$, F the normalized spatial profile with $\iint F dx dy = 1$ and $xyz$ the reference in physical coordinates. The evolution of the slowly varying temporal profile $A$ will be governed by the 1D nonlinear Schrödinger (NLS) integrable model:[14]

$$i\frac{\partial A}{\partial z} - \frac{\beta_2}{2}\frac{\partial^2 A}{\partial T^2} - \gamma |A|^2 A = 0 \qquad (1)$$

where $\gamma$ is the effective temporal nonlinearity, $\beta_2 = \partial^2 \beta / \partial \omega^2$ is the soliton group velocity dispersion (GVD), and $T = t - z/v_g$ is a moving time frame with $t$ the time and $v_g$ the group velocity. Eq. (1) supports temporal bright solitons for $\beta_2<0$ and dark solitons for $\beta_2>0$. [14] In the first case, as the pulses are confined within a stable (2+1)D spatial soliton, the resulting spatio-temporal soliton (in the commonly accepted extension of the term to non-integrable models), corresponds to a train of (3+1)D self-localized packets or "light bullets"; their stability stems from decoupling the (3+1)D problem into a (2+1)D spatial (nonlocal) and a (1+1)D temporal (Kerr) cases, both of which are known to lead to stable solutions. [5, 14, 22] The previous statement relies on assuming that: i) the repetition rate $1/\sigma$ is large enough for the nonlocal nonlinearity to be insensitive to the pulsed nature of the excitation and ii) the peak intensity induces a negligible Kerr self-focusing. The former hypothesis will be justified later on as we address in detail the response of reorientational media. For the latter we recall that -in simplistic terms- a spatial soliton is obtained when self-focusing balances diffraction; we can express the diffraction angle of a Gaussian beam as $\theta_d = 0.61\lambda_0 / n_0 w_0$, while the Kerr self-focusing angle is $\theta_f = \sqrt{2n_2 P_{peak} / \pi n_0 w_0^2}$ with $w_0$ the beam waist, $P_{peak}$ the peak power, $n_0$ and $n_2$ the linear and intensity-dependent indices of refraction, respectively. For $P_{peak}$ much lower than the Kerr spatial soliton critical power, nonlinear diffraction can be approximated by $\theta_d^* = \sqrt{\theta_d^2 - \theta_f^2}$; therefore, for input powers able to excite a nonlocal spatial soliton, the figure $\xi \equiv (\theta_d - \theta_d^*)/\theta_d$ quantifies the impact of the Kerr effect on the spatial evolution. For highly nonlocal media, moderate peak excitations and typical (electronic) $n_2$ values (of the order of $10^{-17}$ m$^2$/W or lower), it is straightforward to estimate $\xi \ll 1$, i. e. Kerr self-focusing can be neglected.

In order to demonstrate the physical impact of the above, we now address the specific case of a nonlocal soliton excited and propagating in nematic liquid crystals (NLC), within a structure similar to that describer in Ref. [6] but with no applied voltage: two parallel glass slides sandwich the fluid NLC dielectric, with anchoring layers aligning the optic axis $\hat{n}$ (molecular director) parallel to the cell plane $yz$, with $z$ the wavevector direction and $y$ orthogonal to the NLC thickness along $x$. The input beam, extraordinarily polarized in $yz$, propagates in the medium at an angle $\theta = \theta_0$ with respect to the director. For waists significantly smaller than the NLC thickness, neglecting vectorial effects and adopting the paraxial approximation, the evolution of the beam spatial profile F(x,y) is given by:

$$2ik\frac{\partial F}{\partial z} + \frac{\partial^2 F}{\partial x^2} + \frac{\partial^2 F}{\partial y^2} + k_0^2 \left[n^2(\theta) - n_0^2(\theta)\right] F = 0 \qquad (2)$$

where $k_0$ is the vacuum wavenumber, $k^2 = k_0^2 \left[n_0^2 + n_a^2 \sin^2(\theta_0)\right]$, $n_a^2 = n_\parallel^2 - n_\perp^2$ the optical anisotropy and $\theta$ the angular distribution of the director $\hat{n}$, governed by [6, 7, 25]:

$$4K\left(\frac{\partial^2 \theta}{\partial x^2}+\frac{\partial^2 \theta}{\partial y^2}\right)+\left(n_\parallel^2-n_\perp^2\right)\sin(2\theta)|F|^2\left\langle|\Sigma_n A(z,T-n\sigma)|^2\right\rangle_t=0 \quad (3)$$

In Eq. (3) $K$ is the NLC elastic constant and $<>_t$ is a time-mobile average operator to account for the slow response of molecular reorientation. In the absence of excitation, the time-evolution of a perturbed $\theta$-distribution is given by:

$$\frac{\eta}{K}\frac{\partial \theta}{\partial t}=\frac{\partial^2 \theta}{\partial x^2}+\frac{\partial^2 \theta}{\partial y^2} \quad (4)$$

with $\eta$ the viscosity. For a (narrow) input beam propagating in the sample mid-plane, we can approximate the transverse angle profile as a Gaussian $\theta=\hat{\theta}(t)exp[-(r/W_N)^2]$. Substituting in Eq. (4), we obtain:

$$\frac{d\hat{\theta}}{dt}=K\frac{4}{\eta}\left(\frac{r^2}{W_N^4}-\frac{1}{W_N^2}\right)\hat{\theta} \quad (5)$$

In the highly nonlocal regime the soliton waist $W_s$ is substantially narrower than the perturbation, i.e. $W_0 \ll W_N$; hence, for $r$ comparable to $W_0$, we can adopt the approximation $R^2/W_N^4 \ll 1/W_N^2$, yielding the closed form solution $\hat{\theta}=\hat{\theta}_0 Exp(-t/\tau)$, where $\tau=\eta W_N^2/4K$ is the relaxation time. Considering the commercial NLC mixture E7, a realistic value for $W_N \approx 20\mu m$ [6], $K\approx 10^{-11}N$ and $\eta=0.05$ Pa s [25], we obtain $\tau>100$ ms; hence, excitations at repetition rates higher than ~0.1 kHz affect only the spatial envelope through the average power, whereas the electronic nonlinearity responds to the peak intensity associated to each pulse. In a Gaussian pulse train, average and peak powers are related by $P_{peak}=P_{ave}\sigma/(\tau_p\sqrt{\pi})$ (here $\tau_p$ is the pulse duration) and can be independently adjusted by varying $w_p$ and the pulse separation $\sigma$.

System (2-3) yields stable solitary solutions that can be solved for numerically, with soliton power $P_S$ and waist $W_0$ related by $P_S = 2\lambda^2 Kcn/(\pi n_a^4 W_0^2)$, where $n$ is the mean refractive index for extraordinary rays ($n \approx 1.55$) and $\lambda$ is the wavelength.[23] Approximate $z$-invariant solitons can be calculated using the beam propagation method coupled with a nonlinear relaxation algorithm to solve Eqs. (2-3). [6] In our calculations we used $\lambda$= 850nm and average powers $P_{ave}$ set between 0.5 and 10 mW. Figure 1 shows the intensity distribution of a nonlocal soliton for a Gaussian excitation with $P_{ave}$=4.1mW and field waist (1/e) of 3.6μm launched in a NLC cell of thickness L=60μm (Fig. 1(a)-(b)), the optics axis set at $\theta_0=\pi/6$. The resulting director distribution (Fig. 1(c)) and optical intensity $<E^2>$ (Fig. 1(d)) over a propagation length of 5mm underline the highly nonlocal nature of the medium. The power dependence of the soliton waist is shown in Fig. 2(a). In order to estimate the waveguide dispersion we expressed the wavelength dependence of both ordinary and extraordinary NLC indices with a single resonance [25]:

$$n_\perp = 1 + \Gamma_\perp \lambda^2/(\lambda^2 - \lambda_\perp^2) \qquad n_\parallel = 1 + \Gamma_\parallel \lambda^2/(\lambda^2 - \lambda_\parallel^2) \qquad (6)$$

where $\Gamma_\perp$, $\Gamma_\parallel$ are the strengths and $\lambda_\perp, \lambda_\parallel$ are the wavelengths of the resonances for $n_\perp$ and $n_\parallel$ respectively. For E7, using a best-fit approach we determined $\lambda_\perp = 129 nm$, $\Gamma_\perp = 0.6751$, $\lambda_\perp = 182 nm$, with a standard deviation $\delta < 5 x 10^{-4}$ in the range 436nm < λ < 1550nm. Furthermore, we evaluated the GVD versus the soliton power by way of a finite-difference mode-solver. We neglected the slight filament ellipticity induced by the boundary conditions as it does not play a significant role when the waist is much smaller than the cell thickness, i.e. $W_0$<<L. As apparent in Fig. 2(b), in the considered geometry and power range the GVD is normal ($\beta_2$>0) even for different anchoring angles $\theta_0$, (which affect the soliton-waveguide GVD through changes in the extraordinary index), as visible in Fig. 2(c). The sign of $\beta_2$, however, can be changed by doping the NLC with a suitable dye, considering an additional weak resonance in Eq. (6), with $\Gamma_{res}$ and $\lambda_{res}$ being its strength and wavelength, respectively. Fig. 2(d) displays the calculated anomalous $\beta_2$ of the 2mW spatial soliton waveguide in the case of a weak resonance ($\Gamma_{res} = 10^{-9}$) for various $\lambda_{res}$.[26] Therefore, the formation of spatio-temporal solitary waves in the form of self-confined light bullets for $\beta_2<0$ can be realistically addressed in such a medium. We will neglect the losses introduced by the doping and the inherent medium scattering as the former depends essentially on the resonance width and can be very small for significant negative $\beta_2$ (lower than -1ps$^2$/m) while the latter depends on the temperature and on the laser wavelength, neither of which are constraints in our approach. Figure 3 shows the calculated evolution of an input pulse $A(z,T) = \sqrt{P_0}\operatorname{sech}(T/T_0)$ (the bright soliton solution of Eq. (1)), with $T_0 = 150$fs, $\beta_2=$ =-0.163ps$^2$/m obtained for a dye resonance at $\lambda_{res} = 1000nm$ and a soliton waveguide excited using an average power $P_{ave}$=3.4mW. The electronic Kerr coefficient for the E7 is $n_2 \approx 5.2 \times 10^{-18}$ m$^2$/W (~10$^{-11}$esu). [27-28]. In the soliton waveguide the nonlinear strength is expressed as $\gamma = n_2\omega_0/cA_{eff}$, with $A_{eff} = \pi W_0^2$ the effective area [16] and $W_0 = 2.5\mu m$ (at the given excitation). As $P_0$ increases from 100mW (Fig. 3(a)) to 9.4W (Fig. 3(b)), temporal confinement takes place, as expected. Noteworthy, the average power can be kept constant by decreasing the repetition rate while the peak power is increased, or by time-chopping a dense pulse train and varying the modulation duty-cycle.

In the case $\beta_2>0$, dark soliton solutions of the 1-D NLS exist and have the form: [16]

$$A(z,T) = \sqrt{P_S}\tanh(T/T_0) \qquad (7)$$

where $T_0$ defines the notch duration and $P_S = \beta_2/(\gamma T_0^2)$ is the peak power of the continuous-wave background. This presence of a CW power pedestal [5, 27] would affect the time-average excitation properties, however, narrow dark solitons can be excited on top of pulses with a large dispersion length compared to the notch waist, [29, 30] enabling the independent control of the average power and the realization of *anti-bullets* under experimentally feasible conditions. An input pulse exciting such a *realistic* fundamental dark soliton is $A(z,T) = \sqrt{P_0}\tanh(T/T_0)\operatorname{Exp}\left[-(T/\tau_p)^2\right]$, with $\tau_p \gg T_0$. A train of broad optical pulses, each containing a π phase-discontinuity at the peak of the narrow notch, can be generated by spectral filtering [29, 30]. Spatial (bright) and temporal (dark) soliton requirements for average and peak powers yield a critical relationship between the repetition rate σ and $\tau_p = \sigma[4c\lambda Kn\, n_2 T_0^2/(\pi^{3/2} n_a^4 w_0^4 \beta_2)]$. Figure 4 shows the calculated evolution of an input pulse of the form Eq. (7), launched on a top of a broad pulse with $\tau_p = 6.5$ps and $T_0 = 150$fs, with average power $P_{ave} = 3.4$mW in both the low ($P_0 = 100$mW) and high ($P_0 = 10$W) peak-power limits, the latter close to the critical value. Clearly, the notch disperses for low $P_0$ (Fig. 4(a)) but maintains its profile when the Kerr response becomes substantial (Fig. 4(b)), in agreement with the results obtained in Ref. [29-30] for fundamental dark-solitons on a finite-background.

In conclusion, we presented a novel approach to spatio-temporal solitons and (3+1)D nonlinear pulse shaping in media where two different nonlinearities define spatial and temporal responses. The synergetic action of nonlocal and instantaneous nonlinearities in space and in time, respectively, can be combined with independently tuned strengths owing to their distinct characteristic scales: such a decoupling enables to "access" spatio-temporal self-localized wave packets under experimentally feasible conditions. We numerically demonstrated self-confined light *bullets* and *anti-bullets*, i.e. bright and dark temporal solitons propagating in bright 2+1D spatial solitons, in reorientational nematic liquid crystals at wavelengths in the visible or near-infrared. These results carry a large potential impact on localized light propagation in media with nonlinear and nonlocal optical responses (e.g., thermo-optic and photorefractive dielectrics) as well as on physical systems where multiple nonlinear mechanisms coexist (e.g. matter waves in fluids, plasmas, Bose-Einstein condensates). Given the variety of non-instantaneous nonlocal dielectric systems identified in recent years, we expect these results to open up new avenues towards the realization of stable spatio-temporal solitons.

**Acknowledgements**  This work is supported by a NSERC Discovery grant. In addition, IBB acknowledges support from NSERC-USRA and NSERC PGS-M fellowships. MP thanks the Marie Curie People action through project TOBIAS PIOF-GA-2008-221262.

Figure Captions

Figure 1 - Numerical integration of system (2)-(3) for an E7 type NLC in a cell of thickness L=60μm. A Gaussian beam of average power $P_{ave}$=4.1mW and a field waist (1/e) of 3.6μm is launched along *z*. (a-b) A (2+1)D spatial soliton forms and propagates with a transverse invariant profile. The distributions of (c) director orientation and (d) intensity $<E^2>$ depicted at *z*=5mm emphasize the nonlocal character of the reorientational response.

Figure 2 – Calculated soliton-waveguide (a) waist $W_0$ and (b) GVD vs. the average power of the pulse train for $\theta_0=\pi/6$ and (c) calculated soliton GVD versus initial director orientation $\theta_0$ for an average power $P_{ave}$=3.4mW. All the plots consider the propagation in undoped NLC λ=850nm. (d) Dependence of the soliton-waveguide GVD on the NLC dye doping resonant wavelength for a strength $\Gamma_{res}=10^{-9}$, $\theta_0=\pi/6$, $P_{ave}$=3.4mW, λ=850nm.

Figure 3 – Temporal evolution of the intensity (normalized units) of the bright pulse for $T_0$ = 150fs and $\beta_2$= =-0.163ps$^2$/m: (a) at low peak-power $P_0$ = 100mW the pulse disperses in propagation whereas (b) at high peak-power $P_0$ = 9.4W a bright soliton forms. The average power of the train is 3.4mW in both cases.

Figure 4 – Temporal evolution of the intensity (normalized units) of the dark pulse on a finite-background for $T_0$ = 150fs and $\tau_p$ = 6.5ps: (a) at low peak-power $P_0$ = 100mW the dark notch disperses in propagation whereas (b) at high peak-power $P_0$ = 10W a dark soliton forms. The average power of the train is 3.4mW in both cases.

Figure 1

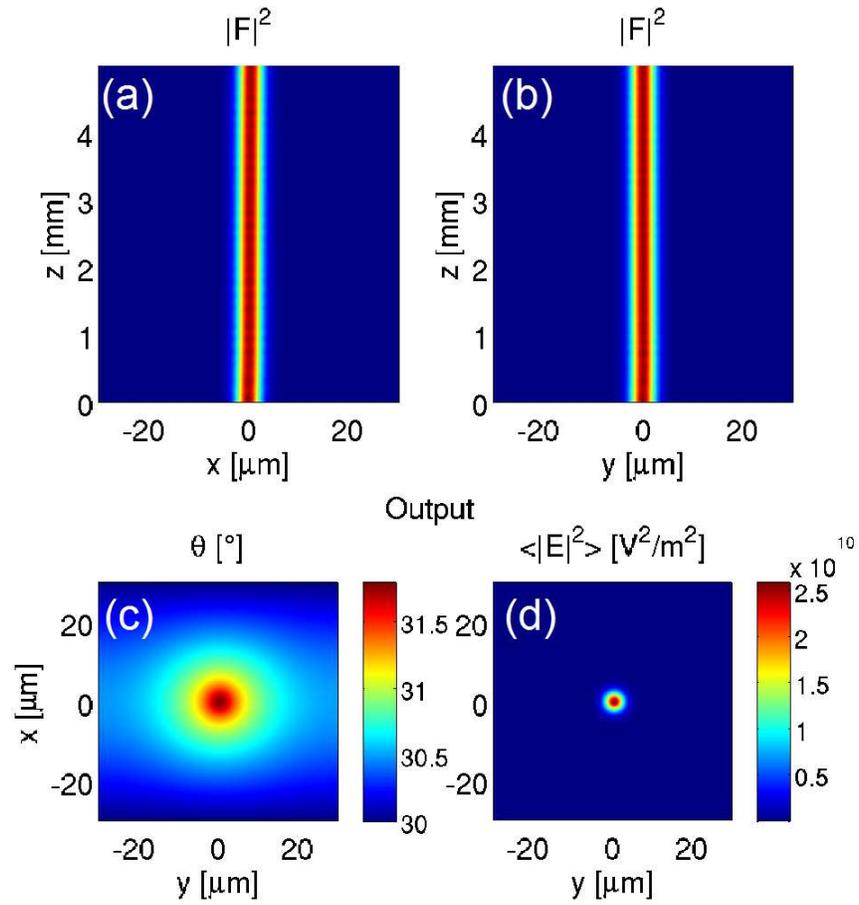

Figure 2

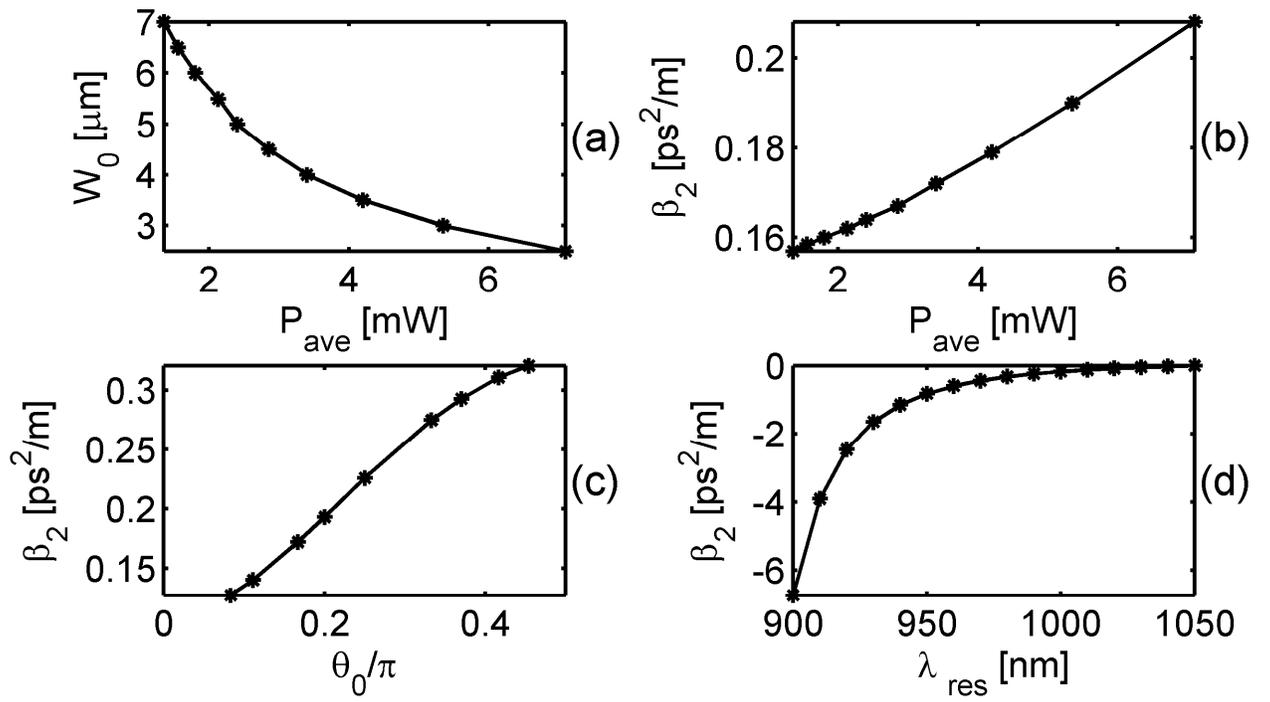

Figure 3

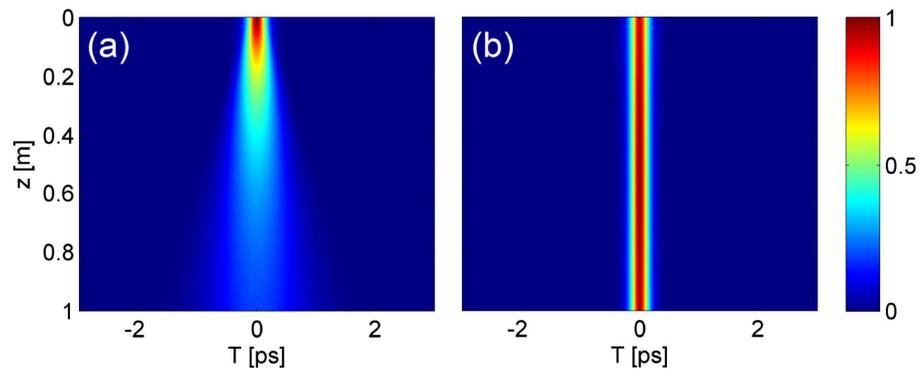

Figure 4

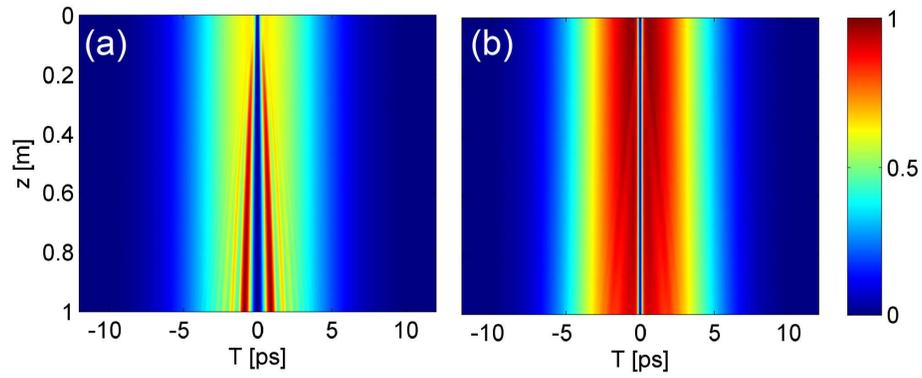

* Corresponding author, email: peccianti@emt.inrs.ca